\documentclass[
aip,
jap,
reprint,
groupedaddress
]
{revtex4-2}

\usepackage[utf8]{inputenc}
\usepackage{siunitx}
\usepackage{graphicx}
\usepackage{amssymb}
\usepackage{booktabs}
\usepackage{color,xcolor}
\usepackage{hyperref}

\newcommand{\lwr}[1]{\textsubscript{\protect\raisebox{-1pt}{#1}}}

\newcommand{\mlwr}[1]{_{\mathrm{#1}}}		
\newcommand{\mupr}[1]{^{\mathrm{#1}}}

\definecolor{Maroon}{cmyk}{0, 0.87, 0.68, 0.32}

\begin{document}

\title[C\lwr{i}-related defect levels in 4$H$-SiC]{Formation of carbon interstitial-related defect levels by thermal injection of carbon into $n$-type 4$H$-SiC}

\author{Robert Karsthof}
\affiliation{Centre for Materials Science and Nanotechnology, University of Oslo, 0316 Oslo, Norway}
\email{r.m.karsthof@smn.uio.no}

\author{Marianne Etzelm\"{u}ller Bathen}
\affiliation{Advanced Power Semiconductor Laboratory, ETH Zürich, Physikstrasse 3, 8092 Zürich, Switzerland}
\altaffiliation{Department of Physics/ Centre for Materials Science and Nanotechnology, University of Oslo, 0316 Oslo, Norway}

\author{Andrej Kuznetsov}
\affiliation{Department of Physics, University of Oslo, 0316 Oslo, Norway}

\author{Lasse Vines}
\affiliation{Department of Physics, University of Oslo, 0316 Oslo, Norway}

\date{September 2021}

\begin{abstract}
        Electrical properties of point defects in 4$H$-SiC have been studied extensively, but those related to carbon interstitials (C\lwr{i}) have remained surprisingly elusive until now. Indeed, when introduced via ion irradiation or implantation, signatures related to C\lwr{i} observed by deep level transient spectroscopy (DLTS) tend to overlap with those of other primary defects, making the direct identification of C\lwr{i}-related levels difficult. Recent literature has suggested to assign the so-called M center, often found in as-irradiated 4$H$-SiC, to charge state transitions of the C\lwr{i} defect in different configurations. In this work, we have introduced excess carbon into low-doped $n$-type \SI{150}{\micro\meter} thick 4$H$-SiC epilayers by thermal annealing, with a pyrolyzed carbon cap on the sample surface acting as a carbon source. Because the layers exhibited initially low concentrations of carbon vacancies  ($[V\lwr{C}] = \SI{e11}{\per\cubic\centi\meter}$), this enabled us to study the case of complete V\lwr{C} annihilation, and formation of defects due to excess carbon, i.e. carbon interstitials C\lwr{i} and their higher-order complexes. We report on the occurrence of several new levels upon C injection which are likely C\lwr{i}-related. Their properties are different from those found for the M center, which points towards a different structural identity of the detected levels. This suggests the existence of a rich variety of C\lwr{i}-related defects. The study will also help generating new insights into the microscopic process of V\lwr{C} annihilation during carbon injection processes.
\end{abstract}

\maketitle

\section{Introduction}

Point defects in the 4$H$ polytype of silicon carbide (4$H$-SiC) exhibit great technological potential in novel quantum technologies; on the other hand, such defects can have detrimental effects on  power device operation. Specifically, the silicon vacancy (V\lwr{Si}) and related complexes can be functionalized for single-photon emission and coherent spin manipulation \cite{Widmann_2014,Castelletto_2014,Christle_2015,von_Bardeleben_2016}. Meanwhile, the carbon vacancy (V\lwr{C}) is notorious for limiting the minority carrier lifetime \cite{Danno2007} and having a detrimental impact on power device performance. However, despite extensive studies on the fundamental defects and their complexes in 4$H$-SiC, the electrical signatures of interstitial defects remain elusive. 

DLTS spectra of as-grown n-type 4$H$-SiC commonly exhibit the ever-present Z\lwr{1/2} and EH\lwr{6/7} levels, which were assigned to the $(0/2-)$ and the deeper-lying (2+/+/0) charge transition levels of the carbon vacancy, respectively \cite{Son2012,Booker2016}. 
Upon irradiation other defect levels appear including the EH\lwr{4} and EH\lwr{5} levels \cite{Storasta2004,Beyer_2012}  (recently tentatively assigned to the carbon antisite-vacancy pair \cite{Karsthof_2020}), the S center \cite{David_2004} (attributed to the Si vacancy \cite{Bathen_3019}), EH\lwr{1/3} \cite{Hemmingsson1997,Alfieri_2020}, and the M center \cite{Nielsen_2003,Martin_2004,Nielsen_2005,Nielsen2005b}.  
Intriguingly, the latter three defect levels (S\lwr{1}/S\lwr{2}, EH\lwr{1}/EH\lwr{3} and M\lwr{1}/M\lwr{3}) are reported in the same regions of the DLTS spectra and often overlap.  
The M center is metastable \cite{Nielsen_2003,Nielsen_2005}, while EH\lwr{1/3} has been found to arise after electron irradiation below the assumed silicon displacement limit \cite{Alfieri_2020}.
Depending on the study, different formation and annealing parameters are reported, and some controversy remains regarding the assignment and nomenclature of the different defect species. 
It was noted that the photoluminescence emission intensity from defects in 4$H$-SiC, e.g. from the V\lwr{Si}, is enhanced after annealing at \SIrange[]{100}{600}{\degreeCelsius} depending on the implantation species \cite{Wang_2019}. This has been partly ascribed to the annealing of C\lwr{i}, a it is expected to be an efficient non-radiative recombination center, reducing the overall luminescence irradtiated samples.


C\lwr{i} was predicted to be electrically active early on \cite{Bockstedte2003,Gali_2003,Kobayashi2019}. Direct identification has proven challenging, but C\lwr{i} may account for instabilities that arise in the deep level transient spectroscopy (DLTS) spectra of n-type 4$H$-SiC immediately after irradiation, but disappear upon heat treatments at \SIrange[]{100}{300}{\degreeCelsius} \cite{Alfieri_2005,Karsthof_2020}. A recent study on He-implanted 4$H$-SiC, combining DLTS measurements and density functional theory calculations,  aimed to address C\lwr{i} directly by assigning 
the M center to the carbon self-interstitial \cite{Coutinho2021}. The bistability was attributed to conversion between C\lwr{i} at the hexagonal (\textit{h}) and pseudo-cubic (\textit{k}) lattice sites. 
However, the unavoidable presence of vacancy-related defects after irradiation prevents the study of interstitial related defects in an unobstructed environment.

The involvement of C\lwr{i} has been discussed extensively in the context of V\lwr{C} control. 
The V\lwr{C} is perennially present in state of the art 4$H$-SiC epitaxial layers \cite{Zippelius2012} and its presence has been correlated to lower minority carrier lifetimes \cite{Danno2007} having a negative impact on device operation. Removal of the V\lwr{C} is thus of strong interest and three main strategies have been devised: (i) near-surface implantation of ions and subsequent annealing leading to V\lwr{C} annihilation \cite{Storasta2008,Hayashi2012,Miyazawa2013,Ayedh2015}, (ii) thermal oxidation of the epilayer surface, promoting injection of Si and C into the bulk material \cite{Hiyoshi2009,Hiyoshi2009a,Miyazawa2013} and (iii) thermal equilibration of V\lwr{C} by annealing in the presence of a carbon cap \cite{Ayedh2015a,Ayedh2017}. Common to all three methods is the presence of an excess of mobile ion species moving through interstitial lattice sites, promoting the disappearance of V\lwr{C} by the ions occupying V\lwr{C} sites. In this regard, carbon appeals as the most suitable ion species since it completely annihilates with V\lwr{C} without introducing new defect levels. While the methods (i)-(iii) have been shown to work well, 
the introduction and migration through the 4$H$-SiC lattice is not well-established because the amount of excess carbon is usually lower than the concentration of V\lwr{C} defects. It is sometimes assumed that diffusion of C mono-interstitials (C\lwr{i}) dominates; however, C\lwr{i} is also known to be a highly unstable defect which that often binds to other defect species, or form carbon complexes. Therefore the question arises whether instead such higher-order complexes of carbon mediate C diffusion. To directly observe C\lwr{i}-related defect levels, epilayers with initially low [V\lwr{C}] are required, such that after C injection and V\lwr{C} annihilation, a surplus of C\lwr{i}-related defects remains.

In this study, we monitor the thermally induced introduction of excess carbon into 4$H$-SiC epi-layers in the near absence of vacancy-related defects. 
Excess carbon is introduced into high-purity and low-doped $n$-type 4$H$-SiC by thermal annealing using  a  pyrolyzed  carbon  cap  on  the  sample surface  acting  as  a  C  source. 
The low concentration of other defects prior to and during annealing enables the separation of interstitial- and vacancy-related defect levels. 
Additionally, the interaction between V\lwr{C} and C\lwr{i} is discussed where we report on several levels appearing upon C injection, with a  structural identity of the detected levels that appears to be different from that of the M center. 

\section{Methods}


The studied 4$H$-SiC samples consisted of \SI{150}{\micro\meter} thick epilayers doped with N ($n \approx \SI{2e14}{\per\cubic\centi\meter}$ as determined by capacitance-voltage measurements) on highly-doped $c$-oriented 2" substrates, purchased from Ascatron AB, Sweden. The samples were laser-cut into pieces of \SI{7x7}{\milli\meter}, cleaned using the full standard RCA procedure, and then coated with photoresist (type AR-U 4030) on the sample front and backside in several spin-coating cycles (total resist thickness $\approx \SI{5}{\micro\meter}$). The coated samples were then heat-treated in an RTP furnace at a temperature of \SI{900}{\degreeCelsius} for \SI{10}{\minute} in vacuum ($p = \SI{5e-5}{mbar}$) during which the photoresist was graphitized. The C-capped samples were then annealed in a tube furnace under Ar atmosphere at \SI{1250}{\degreeCelsius} for \SIlist[list-final-separator={ \text{or} }]{1.5;2.5;6.6}{\hour}. It should be noted that after the high-temperature anneals, the C cap on the samples had been almost completely dissolved, and the samples' surfaces had been oxidized to form SiO\lwr{2} (as determined using energy-dispersive X-ray spectroscopy). We attribute this to the presence of residual oxygen during the annealing process, likely due to continued outgassing of the furnace tube. It is unknown how long into the annealing the C-caps were still intact to provide a source of C. However, thermal oxidation of SiC is another common procedure to inject C into the epilayer; therefore, we believe this to be negligible for the interpretation of this experiment. 

After annealing, the samples were RCA-cleaned and \SI{150}{\nano\meter} thick Ni Schottky diodes were deposited on the epilayer using e-beam evaporation through a shadow mask (diode area \SI{7.85e-3}{\square\centi\meter}). Silver paste was used as a back contact to the conductive substrate. We note that the oxidation of the epilayer surface during the annealing did not lead to a significant surface roughening, nor to a degradation of the Schottky diode quality. 

The electrical characterization (CV) was carried out using a Boonton-7200 high-precision capacitance meter and an Agilent 81110A pulse generator (for DLTS measurements). The DLTS measurements were performed in a temperature range between \SI{20}{\kelvin} and \SI{300}{\kelvin}, where cooling was achieved with a closed-cycle He refrigerator.

\section{Results}

\begin{figure}
    \centering
    \includegraphics[width=\columnwidth]{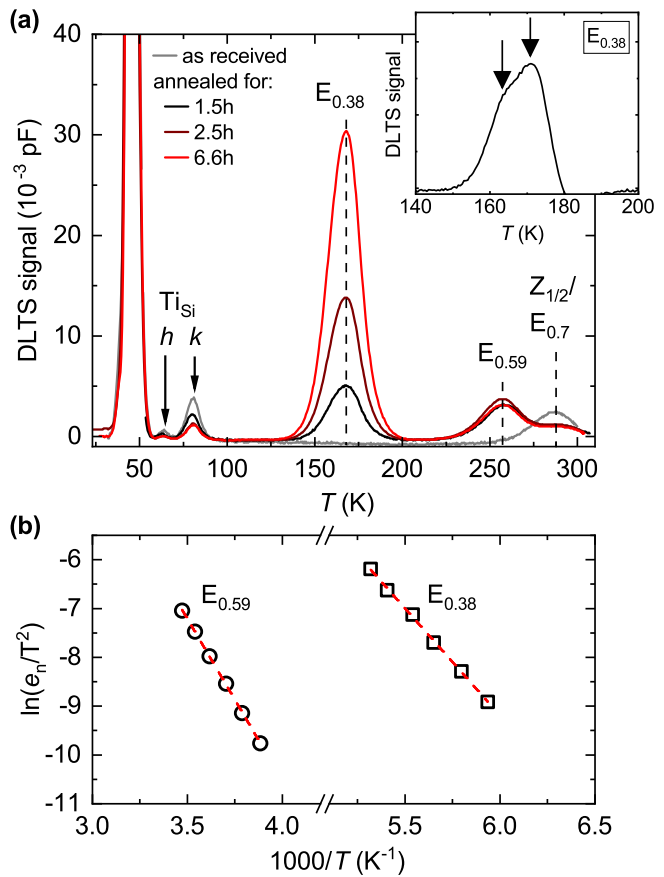}
    \caption{(a) Complete DLTS spectra (rate window $(\SI{640}{\second})\mupr{-1}$) of the as received as well as samples annealed for different durations. Inset: DLTS spectrum of the E\lwr{0.38} trap with a high-resolution weighting function obtained on the \SI{2.5}{\hour} annealed sample, revealing two components of the level. (b) Plot of the reduced emission rates of the E\lwr{0.38} and E\lwr{0.59} traps and respective Arrhenius fits.}
    \label{fig:DLTS_all_samples_Arrhenius_E038_E059}
\end{figure}

In Fig.~\ref{fig:DLTS_all_samples_Arrhenius_E038_E059}(a), DLTS spectra of the as-received samples as well as samples annealed at \SI{1250}{\degreeCelsius} for durations of \SIlist{1.5;2.5;6.6}{\hour} are shown. In the as-received state, peaks belonging to the donor freeze-out (at \SI{50}{\kelvin}), the Ti\lwr{Si} defect, and the V\lwr{C} (Z\lwr{1/2} level) can be detected. The concentration of the V\lwr{C} amounts to \SI{1.5e11}{\per\cubic\centi\meter}. 
After annealing under the C-rich environment of the C-cap, several observations can be made. Most prominently, two new traps appear at temperatures of about \SI{170}{\kelvin} (labeled E\lwr{0.38}) and \SI{265}{\kelvin} (labeled E\lwr{0.59}), respectively. Their activation energies and apparent electron capture cross sections, determined from the reduced emission rate data in Fig.~\ref{fig:DLTS_all_samples_Arrhenius_E038_E059}(b), are given in Table~\ref{tab:exp_trap_prop}. Note that the activation energy and apparent electron capture cross section for the \SI{170}{\kelvin} peak are similar to the values previously found on the M\lwr{1} level ($E\mlwr{a} = \SI{0.42}{\electronvolt}$, $\sigma\mlwr{n} = \SI{6e-15}{\square\centi\meter}$) \cite{Nielsen_2005}. 
Given the fact that the M center has previously been shown to originate from charge transition levels (CTLs) of the C\lwr{i} \cite{Coutinho2021}, and taking into account that during the experiment presented in the present manuscript considerable amounts of carbon can be expected to be continuously driven into the uppermost regions of the epilayer, which suggests the presence of the C\lwr{i} defect, it is tempting to identify the trap emitting at \SI{170}{\kelvin} with the M\lwr{1} level. However, the data presented in the further course of this manuscript casts doubt on this identification; therefore, we choose to adhere to the temporary label E\lwr{0.38} within this work. 
Note that while the amplitude of E\lwr{0.38} increases with annealing time, that of the E\lwr{0.59} level is approximately the same for all annealed samples, suggesting a different microscopic origin of the defects behind the two signatures. Although the activation energy of E\lwr{0.59} matches the value predicted for the M\lwr{2} level by DFT, 
it seems unlikely that they are identical, given their different peak emission temperatures (\SI{265}{\kelvin} for E\lwr{0.59} vs. typically \SI{295}{\kelvin} for M\lwr{2}).

A second important observation from Fig.~\ref{fig:DLTS_all_samples_Arrhenius_E038_E059}(a) is that the concentration of the Z\lwr{1/2} level, originating from the $(0/2-)$ CTL of the V\lwr{C} defect, is apparently reduced from an initial value of \SI{3e11}{\per\cubic\centi\meter} (corresponds to a V\lwr{C} concentration of \SI{1.5e11}{\per\cubic\centi\meter}) by roughly a factor of 2 for an annealing time of \SI{1.5}{\hour} and stays approximately constant for longer annealing. Because it is located on the decreasing flank of the E\lwr{0.59} peak, and also occurs nearly outside of the accessible temperature range, the activation energy of the shoulder appearing on the annealed samples can only be estimated by simulations of the DLTS spectra, assuming certain values for $E\mlwr{a}$ and $\sigma\mlwr{n}$. According to such simulations, $E\mlwr{a}$ amounts to roughly \SI{0.7}{\electronvolt}, and is therefore similar to that of Z\lwr{1/2} ($E\lwr{a} = \SI{0.67}{\electronvolt}$). However, the depth profile for this level which is presented further below is incompatible with that expected for Z\lwr{1/2}. We therefore conclude that this level does not originate from V\lwr{C} but from a different structural defect. In the further course of this paper, this level will be labelled E\lwr{0.7}, based on its approximate activation energy. 
We note that the reduction of [Z\lwr{1/2}]  (and therefore [V\lwr{C}]) to below the detection limit of DLTS by the presented annealing procedure is striking and supports the claim of carbon  being injected into the epilayer, leading to C\lwr{i} and V\lwr{C} recombining. 

\begin{table}[]
    \centering
        \caption{Experimentally determined properties (activation energy $E\mlwr{a}$, apparent electron capture cross section $\sigma\mlwr{n,app}$ and peak temperature T) of the E\lwr{0.38}, E\lwr{0.59}, and E\lwr{0.7} traps.}
    \begin{tabular}{cccc}
        \toprule
        trap label & {$E\mlwr{a}$ (\si{\electronvolt})} & {$\sigma\mlwr{n,app}$ (\si{\square\centi\meter})} & T (\si{\kelvin}) \\
        \midrule
         E\lwr{0.38} & \num{0.38} & \num{5e-15} & 170 \\
         E\lwr{0.59} &  \num{0.59} & \num{2e-14} & 265 \\
         E\lwr{0.7} & \num{0.7} & \num{4e-13} & 295\\
         \bottomrule
    \end{tabular}
    \label{tab:exp_trap_prop}
\end{table}

\begin{figure}
    \centering
    \includegraphics[width=\columnwidth]{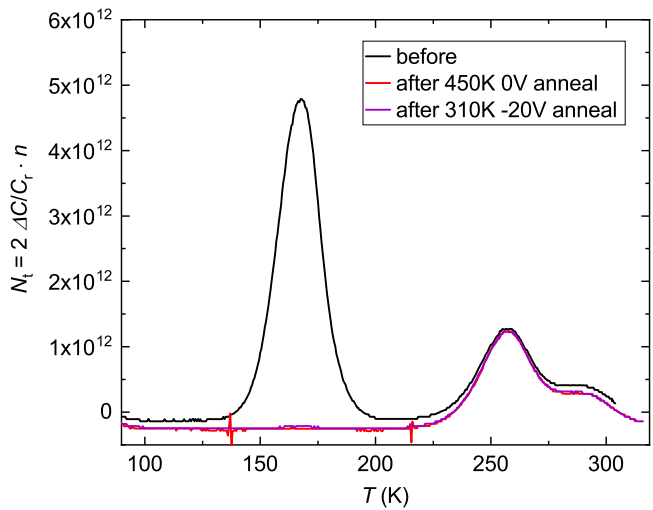}
    \caption{Bias annealing behavior of the defect signatures detected by DLTS upon C injection, collected on the \SI{2.5}{\hour}-annealed sample.}
    \label{fig:DLTS_bias_annealing}
\end{figure}

A characteristic behavior of the M center is its bistability with respect to annealing at moderate temperatures (up to \SI{200}{\degreeCelsius}) with and without applied bias \cite{Nielsen2005b,Beyer.2011,Capan.2021}. Specifically, an annealing at zero bias and $T = \SI{450}{\kelvin}$ leads to the reduction of the M\lwr{1} (and M\lwr{3} which lies outside of the studied range) amplitude and the emergence of the M\lwr{2} level; annealing at \SI{310}{\kelvin} with large reverse bias will restore the original DLTS spectrum without loss of the M\lwr{1} and M\lwr{3} amplitudes. The specified annealing procedures (reverse bias annealing at $V = \SI{-20}{\volt}$ for \SI{20}{\minute}; zero-bias anneals at \SI{0}{\volt} for \SI{20}{\minute}) have been performed on the \SI{2.5}{\hour}-annealed sample, the results of which are shown in Fig.~\ref{fig:DLTS_bias_annealing}. As can be seen in the figure, the E\lwr{0.38} amplitude is reduced to below the detectivity limit ($<\SI{5e10}{\per\cubic\centi\meter}$) by zero-bias annealing at \SI{450}{\kelvin}; however, it remains undetectable after the reverse annealing step which is incompatible with E\lwr{0.38} being identical to M\lwr{1}. Moreover, the amplitudes of E\lwr{0.59} and E\lwr{0.7} remain unaffected by the annealing procedure, supporting the hypothesis that they originate from another structural defect -- one that is more stable than E\lwr{0.38}. Importantly, they are likewise unrelated to the M center.

\begin{figure}
    \centering
    \includegraphics[width=\columnwidth]{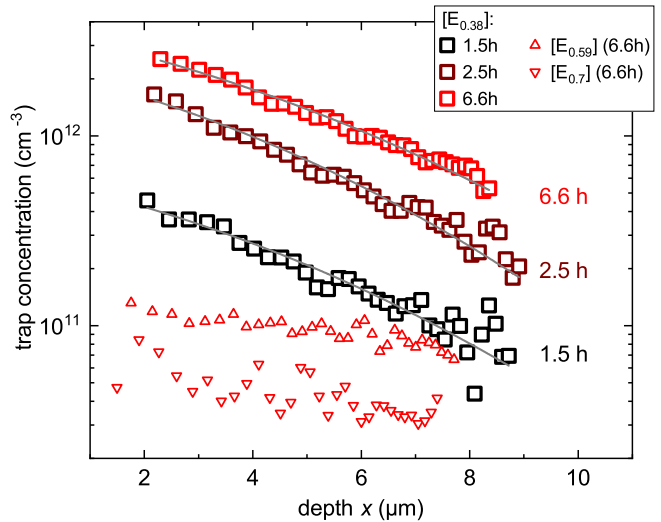}
    \caption{Concentration profiles of the E\lwr{0.38}, E\lwr{0.59} and E\lwr{0.7} levels, determined by DLTS profiling. Profiles of [E\lwr{0.38}] are shown for all annealing times. Profiles of  [E\lwr{0.59}] and [E\lwr{0.7}] are shown only for the \SI{6.6}{\hour} anneal. Fits to the E\lwr{0.38} profiles according to Eq.~\ref{eq:diffusion_conc} are shown in gray.}
    \label{fig:DLTS_profiles_all_traps}
\end{figure}

Concentration versus depth profiles for the traps
E\lwr{0.38}, E\lwr{0.59}, and E\lwr{0.7} were recorded by varying the
voltage pulse height, from $V\mlwr{p} = \SI{0.5}{\volt}$ to \SI{20}{\volt}, using a reverse bias of \SI{-20}{\volt}, and at a fixed temperature.  
The $\lambda$ correction was taken into account. The resulting trap concentration profiles are displayed in  Fig.~\ref{fig:DLTS_profiles_all_traps}. For E\lwr{0.38}, a clear increase in concentration towards shallower depths is seen, in compliance with a defect entering the epilayer from the surface. For the E\lwr{0.59} and E\lwr{0.7} levels, the signal amplitude is lower than for E\lwr{0.38}, however a slight decrease towards the sample bulk can be observed here as well. The E\lwr{0.38} concentration profiles have been fitted with the relation

\begin{equation}
    c(x) = \frac{c_0}{2} \cdot \mathrm{erfc}\left\{\frac{x}{2\sqrt{Dt\mlwr{ann}}} \right\}
    \label{eq:diffusion_conc}
\end{equation}

where $c\mlwr{0}$ is the concentration of E\lwr{0.38} at $x = 0$, $D$ is the diffusion constant, and $t\mlwr{ann}$ the annealing time. From these fits to the concentration profiles obtained from DLTS for $t = \SI{1.5}{\hour}$, \SI{2.5}{\hour}, and \SI{6.6}{\hour}, $D$ and $c_0$ were determined and are given in Table~\ref{tab:diffusion}. The diffusion is expected to be thermally activated according to 

\begin{equation}
    D(T) = D\mlwr{0} \exp\left\{- \frac{E\mlwr{i}}{k\mlwr{B}T} \right\}
    \label{eq:diffusion_const}
\end{equation}

with $D_0$ being the pre-exponential factor, usually $D_0 \approx \SI{e-3}{\square\centi\meter\per\second}$ for atomic hops, and $E\mlwr{i}$ the injection barrier. An estimate of $E\mlwr{i}$ can be given based on Eq.~\ref{eq:diffusion_const}, and the values are also given in Table~\ref{tab:diffusion}. Although there is some variation between the different $t\mlwr{ann}$, it is evident that the injection barrier is around \SI{2.35}{\electronvolt}, in accordance with values predicted by DFT for the diffusion of the neutrally charged C\lwr{i} defect along the $c$ axis of 4$H$-SiC ($E\mlwr{m,th} = \SI{2.20}{\electronvolt}$) \cite{Coutinho2021}. We therefore refer to the injection barrier for E\lwr{0.38} as the migration barrier for C\lwr{i} for the remainder of this paper, although it is possible that an additional barrier for release from the C cap is incorporated in this value. 

The depth profiles of E\lwr{0.59} and E\lwr{0.7} appear to be flatter than that of E\lwr{0.38}, but retain a slight downward slope towards the sample bulk. The low concentrations and high noise levels prevent a reliable extraction of formation barriers for the two deeper levels (the detection limit for defects under the conditions of the experiment is approximately \SI{5e10}{\per\cubic\centi\meter}). We observe that although E\lwr{0.59} and E\lwr{0.7} also appear to be coming in from the surface they are likely of a different origin than E\lwr{0.38}. Importantly, the depth profiles in Fig.~\ref{fig:DLTS_profiles_all_traps} emphasize that the E\lwr{0.7} level is likely unrelated to V\lwr{C} because in that case, as was noted further above, its concentration should decrease towards the surface as a result of the C injection.

\begin{table}[]
    \centering
    \caption{Diffusion parameters of the E\lwr{0.38} trap, determined from DLTS depth profiles in Fig.~\ref{fig:DLTS_profiles_all_traps}: diffusion constant $D$, surface concentration $c_0$, and injection barrier $E\mlwr{i}$.}
    \begin{tabular}{SSSS}
    \toprule
        {$t\mlwr{ann}$ (\si{\hour})} & {$D$ (\si{\square\centi\meter\per\second})} & {$c_0$ (\si{\per\cubic\centi\meter})} & {$E\mlwr{i}$ (\si{\electronvolt})} \\
    \midrule
        1.5 & \num{2.65e-11} & \num{1.2e12} & \num{2.28} \\
        2.5 & \num{1.41e-11} & \num{4.6e12} & \num{2.37} \\
        6.6 & \num{6.85e-12} & \num{7.3e12} & \num{2.46} \\
    \midrule
        {avg.} & & & \num{2.35} \\
    \bottomrule
    \end{tabular}
    \label{tab:diffusion}
\end{table}

\section{Discussion}

\begin{figure}
    \centering
    \includegraphics[width=\columnwidth]{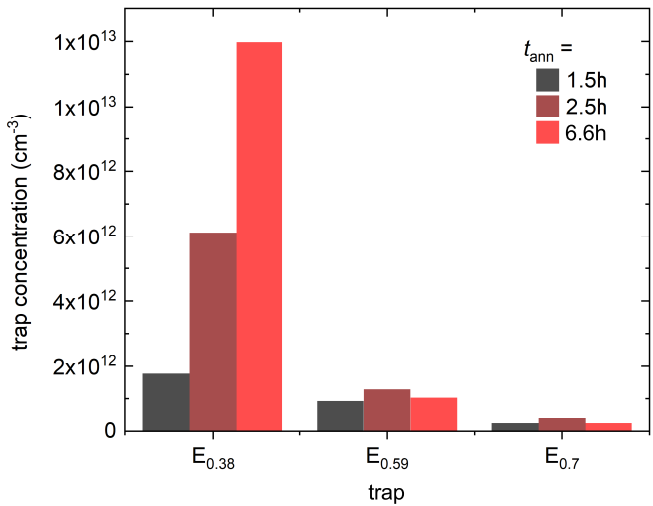}
    \caption{Concentrations of all four traps induced by annealing with C-cap in dependence on annealing time.}
    \label{fig:trap_conc_all}
\end{figure}

In Fig.~\ref{fig:trap_conc_all}, the evolution of the concentrations of the three annealing-induced traps is shown for varying $t\mlwr{ann}$. It can be seen that while  [E\lwr{0.38}] clearly increases with longer anneals, [E\lwr{0.59}] and [E\lwr{0.7}] do not seem to show such a systematic behavior. It therefore appears that, while all three traps are formed by C injection, they originate from structurally different defects. This is also supported by their different annealing behavior (see Fig.~\ref{fig:DLTS_bias_annealing}). As stated already, the injection barrier for the E\lwr{0.38} level matches well with the predicted value for C\lwr{i} migration, and its activation energy of \SI{0.38}{\electronvolt} is close to that of the M\lwr{1} level which has previously been tied to the {$(-/2-)$} acceptor level of the C\lwr{i} \cite{Coutinho2021,Capan.2021}. However, the M center possesses four charge transition levels in total, labelled M\lwr{1} to M\lwr{4}. They appear in pairs (M\lwr{1} and M\lwr{3}, M\lwr{2} and M\lwr{4}) which are now believed to belong to two configurations of the carbon split interstitial that can be reversibly converted into each other. M\lwr{1,2} are thought to originate from the $(-/2-)$ CTL, while M\lwr{3,4} stem from the deeper-lying $(0/-)$ CTL. Depending on the orientation of the C-C dimer, one of the two level pairs becomes dominant over the other, and because the conversion between them is connected to barriers $< \SI{2}{\electronvolt}$, it can be triggered repeatedly by bias annealing procedures close to room temperature. In the experiments for the present paper, however, there were no further signatures resembling those of the M center apart from E\lwr{0.38} which matches M\lwr{1}; DLTS scans up to \SI{360}{\kelvin} revealed no further peaks being present. Taking into account the lack of the reversible character of the bias annealing of the observed defect signatures, as reported for the M center, we therefore suggest that the E\lwr{0.38} level is different from M\lwr{1}, although it is most likely also related to the C\lwr{i} defect. 

E\lwr{0.59} and E\lwr{0.7} do not exhibit a monotonous variation of their concentration with C injection time; however, their occurrence only after the annealing experiment, and the possible slight increase of their concentration profile towards the sample's surface (see Fig.~\ref{fig:DLTS_profiles_all_traps}), suggest that they originate from a defect that has been driven into the SiC epilayer during the anneals. Okuda \textit{et al.} have reported that the thermal oxidation of 4$H$-SiC followed by an anneal in Ar atmosphere at \SI{1550}{\degreeCelsius} leads to the formation of a pair of electron traps labelled ON0a and ON0b \cite{Okuda.2015}. The authors propose that ON0a is identical to the EO1 minority trap found in $p$-type material after thermal oxidation, hence a correlation with carbon injection is conceivable. ON0a and EO1 possess the same energy with respect to the conduction band edge ($E\mlwr{t} = E\mlwr{C} - \SI{0.59}{\electronvolt}$) which also coincides with the activation energy of the E\lwr{0.59} trap found in this work. Okuda \textit{et al.} do not report the activation energy of ON0b; however, its maximum emission occurs at a temperature roughly \SI{30}{\kelvin} higher than for ON0a, which is the same temperature difference as between the E\lwr{0.59} and E\lwr{0.7} traps in the present work. It seems therefore likely that the E\lwr{0.59} and E\lwr{0.7} levels are the same as those labeled ON0a and ON0b by Okuda \textit{et al.}, respectively. Furthermore, the EO1 minority trap has been tentatively attributed to the carbon split di-interstitial (C\lwr{sp})\lwr{2} by Okuda \textit{et al.}, based on ab initio-calculations by Bockstedte, Mattausch and Pankratov \cite{Bockstedte2004}. Specifically, the latter authors calculated the $(0/2-)$ CTLs of the $hh$ and $kk$ configurations of that defect, which is predicted to show negative-$U$ properties, to be situated around \SI{0.55}{\electronvolt} and \SI{0.63}{\electronvolt} below the conduction band minimum, respectively, which roughly matches the measured activation energies for E\lwr{0.59} and E\lwr{0.7}. B, M \& P also calculate the C split di-interstitial to be particularly stable: its predicted dissociation energy amounts to \SI{4.6}{\electronvolt} in both $hh$ and $kk$ configurations, in accordance with the observed temperature stability of E\lwr{0.59} and E\lwr{0.7} as compared to that of E\lwr{0.38}.
An interesting question is how the defect behind E\lwr{0.38} can form at temperatures as high as \SI{1250}{\degreeCelsius}, as well as survive during the cooldown ramp to room temperature (about \SI{6}{\hour} in total), but anneal out at only \SI{450}{\kelvin} during the bias anneals. A possible explanation is that the presence of the built-in electric field of the Schottky barrier promotes the diffusion of C\lwr{i} out of the depletion region. Furthermore, the migration or injection barrier of \SI{2.4}{\electronvolt} determined from the depth profile of E\lwr{0.38} only reflects the high-temperature conditions present during the carbon injection, which promote a mid-gap Fermi level. These typically induce more positive charge states on point defects for which migration barriers are typically higher than when the defect is negatively charged.

Unfortunately, the authors of Ref.~\cite{Bockstedte2004}  did not calculate CTLs of the single C\lwr{i} in 4$H$-SiC to compare with those of di-interstitials. Kobayashi \textit{et al.} have used hybrid-functional DFT calculations to predict the formation energies and CTLs of various native point defects and clusters in 4$H$-SiC \cite{Kobayashi2019}. They find the $(0/-)$ CTL of the carbon split mono-interstitial C\lwr{i,split} in its $k$ and $h$ configurations at $E\mlwr{C} - \SI{0.7}{\electronvolt}$ and $E\mlwr{C} - \SI{0.33}{\electronvolt}$. While the latter is a reasonably good match for E\lwr{0.38} and the former for E\lwr{0.7}, the different injection and annealing behavior for the two traps found in this work contradicts a common structural origin. Kobayashi \textit{et al.} also discuss the case of carbon clusters to a limited extent, and claim that according to their calculations, poly-interstitials exhibit too high formation energies to be stable. Instead, they find the di-carbon antisites (C\lwr{2})\lwr{Si} to be highly stable, and to have their $(0/-)$ CTLs at $E\lwr{C} - \SI{0.5}{\electronvolt}$ ($k$) and $E\lwr{C} - \SI{0.13}{\electronvolt}$ ($h$). Again, while the former approximately matches the activation energy of E\lwr{0.59}, the latter does not seem to have a corresponding trap found in our data (apart from the unequivocally identified Ti\lwr{Si}($h$) level).

Overall, it is challenging to make assignments of the observed trap signatures to structural defects at this point. However, it can be noted that carbon injection, in the near absence of V\lwr{C}, leads to the formation of several new defect levels related to C excess, and that there are at least two physically differing defect structures involved: a less stable one, increasing in concentration as more carbon is injected, possibly hinting at a mono-interstitial as its structural origin, and a more stable one that seems to possess a more complicated formation and annealing behavior, which may be due to a larger carbon complex. Annihilation of V\lwr{C} by carbon injection therefore seems to be a rather complicated process, involving the migration of C in the form of several defect species. It must also be stated that the absence of the M center in this work demonstrates that excess carbon forms a larger variety of interstitial-related defects than previously thought.


\section{Conclusion}

Using a carbon cap on the surface of $n$-type 4$H$-SiC epilayers, we have injected excess carbon by thermal annealing. Because the initial concentrations of carbon vacancies V\lwr{C} was low, the C injection lead to (1) the complete annihilation of V\lwr{C}, and (2) the formation of new C excess-related defects. A set of three defect levels becomes visible in low-temperature DLTS spectra, with activation energies of \SIlist{0.38;0.59;0.7}{\electronvolt}. Depth profiles of the trap levels show that they originate from C injection (the defect concentration increases towards the epilayer surface), and the activation energy associated with the evolution of the depth profiles matches the migration energy for carbon diffusion previously calculated by \textit{ab initio} methods. The dissimilar behavior of the traps with longer injection duration as well as their starkly differing thermal stability points towards a different defect structures behind the signatures, albeit likely related to carbon interstitials in all cases. An identification of either of the defects with the M center, which has previously been attributed to the mono-interstitial C\lwr{i}, does not appear justified.  However, the ON0a and ON0b defects found in thermally oxidized material, and possibly originating from the carbon split di-interstitial (C\lwr{sp})\lwr{2}, match the levels at $E\mlwr{C} - \SI{0.59}{\electronvolt}$ and $E\mlwr{C} - \SI{0.7}{\electronvolt}$ well. Generally, this work emphasizes the complexity of the V\lwr{C} annihilation process via supply of excess carbon, which seems to take place through a variety of channels.

\section*{Acknowledgments}
Financial support was kindly provided by the Research Council of Norway and the University of Oslo through the frontier research project FUNDAMeNT (no. 251131, FriPro ToppForsk-program). 
The Research Council of Norway is acknowledged for the support to the Norwegian Micro- and Nano-Fabrication Facility, NorFab, project number 295864.
The work of MEB was supported by an ETH Zurich Postdoctoral Fellowship. 

\section*{DATA AVAILABILITY}
The data that support the findings of this study are available from the corresponding author upon reasonable request. 

\section*{}
This article has been submitted to Journal of Applied Physics. After it is published, it will be found at \href{https://aip.scitation.org/journal/jap}{link}.

\section*{References} 
\bibliography{refs}

\end{document}